\documentclass[pss,fleqn]{w-art}
\usepackage{times}
\usepackage{graphicx}
\begin{document}
\DOIsuffix{theDOIsuffix}
\Volume{XX}
\Issue{1}
\Month{01}
\Year{2005}
\pagespan{1}{}
\Receiveddate{1 October 2005}
\Reviseddate{}
\Accepteddate{}
\Dateposted{}

\title[The CONQUEST code]{Recent progress with large-scale ab initio calculations: the CONQUEST code}

\author[D. R. Bowler]{D. R. Bowler\footnote{e-mail: {\sf david.bowler@ucl.ac.uk}}\inst{1,2}}
\address[\inst{1}]{Dept. of Physics and Astronomy, University College London, Gower Street, London WC1E~6BT, UK}
\author[R. Choudhury]{R. Choudhury\inst{1}}
\author[M. J. Gillan]{M. J. Gillan\footnote{e-mail: {\sf m.gillan@ucl.ac.uk}}\inst{1}}
\author[T. Miyazaki]{T. Miyazaki\footnote{e-mail: {\sf miyazaki.tsuyoshi@nims.go.jp}}\inst{2}}
\address[\inst{2}]{National Institute for Materials Science, 1-2-1 Sengen, Tsukuba, Ibaraki 305-0047, Japan}

\begin{abstract}
  While the success of density functional theory (DFT) has led to its
  use in a wide variety of fields such as physics, chemistry,
  materials science and biochemistry, it has long been recognised that
  conventional methods are very inefficient for large complex systems,
  because the memory requirements scale as $N^2$ and the cpu
  requirements as $N^3$ (where $N$ is the number of atoms).  The
  principles necessary to develop methods with linear scaling of the
  cpu and memory requirements with system size ($\mathcal{O}(N)$
  methods) have been established for more than ten years, but only
  recently have practical codes showing this scaling for DFT started
  to appear.  We report recent progress in the development of the
  \textsc{Conquest} code, which performs $\mathcal{O}(N)$ DFT
  calculations on parallel computers, and has a demonstrated ability
  to handle systems of over 10,000 atoms. The code can be run at
  different levels of precision, ranging from empirical tight-binding,
  through \textit{ab initio} tight-binding, to full \textit{ab
    initio}, and techniques for calculating ionic forces in a
  consistent way at all levels of precision will be presented.
  Illustrations are given of practical \textsc{Conquest} calculations
  in the strained Ge/Si(001) system.
\end{abstract}
\maketitle

\section{Introduction}
\label{sec:introduction}

This paper aims to summarize recent progress in techniques for
performing \textit{ab initio} calculations with computational effort
scaling linearly with system size.  The target systems are intended to
be very large (tens of thousands or hundreds of thousands of atoms),
with the implementation using methods based on density-functional
theory (DFT) and pseudopotentials~\cite{Martin2004}.  The
\textsc{Conquest} DFT
code\cite{hernandez95,hernandez96,goringe97,bowler99,bowler00,bowler01,bowler02}
is designed to perform this kind of calculation, and its
$\mathcal{O}(N)$ capabilities have been tested on systems of up to
16,000 atoms\cite{bowler01}, but up to now it has been limited to
rather simple systems.  We report here the developments which now make
it possible to do $\mathcal{O}(N)$ DFT calculations on non-trivial materials
problems using the \textsc{Conquest} code.

The Car-Parrinello paper of 1985~\cite{Car1985} set a completely new
agenda for computational condensed matter science. In that paper, the
strategy based on density functional theory (DFT), pseudopotentials
and plane waves was formulated in a new and powerful way that
eventually led to its current widespread application across all the
disciplines that are based on a molecular view of matter. Yet soon
after 1985 an important limitation of the strategy had been
recognised. In several papers in the early
1990's~\cite{Baroni1992,Galli1992,kim95,Ordejon1995}, it was pointed
out that the number of computer operations demanded by the
Car-Parrinello strategy would increase ultimately as $N^3$ with the
number of atoms $N$ in the system. At the same time, it was realised
that such a rapid increase must be avoidable, since the locality of
quantum coherence~\cite{Kohn-1959,kohn96} (known as the principle of
near-sightedness\cite{kohn96}) should make it possible
to determine the electronic ground state of a system with a number of
operations that increases only linearly with $N$. Ideas for
reformulating the strategy to achieve $\mathcal{O}(N)$ scaling were proposed at
that time~\cite{Galli1992,li93,mauri93,kim95,Ordejon1995,kohn96,hernandez96}
, and rapidly led to practical $\mathcal{O}(N)$ methods within tight-binding
schemes. The feasibility of
performing $\mathcal{O}(N)$ DFT calculations on systems of many thousands
of atoms was also demonstrated about 8 years ago\cite{goringe97}.  However, the early promise of those ideas has taken a long
time to be turned into practical techniques for performing $\mathcal{O}(N)$ DFT
calculations on systems containing many thousands of atoms. We report
here encouraging new results from the $\mathcal{O}(N)$ \textsc{Conquest}
code~\cite{hernandez96,hernandez95,bowler02} on systems of
nearly 23,000 atoms, which, together with other recent progress,
suggest that the promise is at last being realised.

%
%
%

We start by recalling the principles on which \textsc{Conquest} and a
number of other $\mathcal{O}(N)$ codes (such as \textsc{onetep}, OpenMX and
\textsc{siesta}) are
based~\cite{hernandez96,hernandez95,bowler02,ordejon96,soler02,ozaki01,Ozaki2003,haynes99,gan01,Skylaris2002,Mostofi2002,Mostofi2003,Skylaris2005}.  The locality of
quantum coherence is expressed by the decay of the Kohn-Sham density
matrix $\rho ( {\bf r} , {\bf r}^\prime ) \rightarrow 0$ as $| {\bf r}
- {\bf r}^\prime | \rightarrow \infty$~\cite{Kohn-1959,kohn96}. One
way of determining the self-consistent DFT ground state is to write
the total energy $E_{\rm tot}$ in terms of $\rho ( {\bf r} , {\bf
  r}^\prime )$, and to minimise $E_{\rm tot}$ with respect to $\rho$,
subject to the conditions that (i)~$\rho$ is Hermitian; (ii)~$\rho$ is
idempotent (i.e. it is a projector); and (iii)~$\rho$ gives the
correct number of electrons~\cite{hernandez96,hernandez95}.  Since
$E_{\rm tot}$ is variational, $\mathcal{O}(N)$ operation is achieved by imposing
the truncation $\rho ( {\bf r} , {\bf r}^\prime ) = 0$ for $| {\bf r}
- {\bf r}^\prime | > r_{\rm c}$, and the true DFT ground state is
recovered as the cut-off distance $r_{\rm c}$ is increased.

To obtain a workable scheme, it is required that $\rho$ be separable
(i.e. that the number of its non-zero eigenvalues be finite):
\begin{equation}
\rho ( {\bf r} , {\bf r}^\prime ) = \sum_{i \alpha , j \beta}
\phi_{i \alpha} ( {\bf r} ) K_{i \alpha , j \beta}
\phi_{j \beta} ( {\bf r}^\prime ) \; .
\end{equation}
In practical codes, the $\phi_{i \alpha} ( {\bf r} )$ have been chosen
to be ``localised orbitals'' centred on the atoms, with $\phi_{i
  \alpha}$ being the $\alpha$th orbital on atom $i$.  (Centring on
atoms is not essential, and the orbitals can be allowed to float, see
for example ~\cite{Fattebert2004}.)  The localised orbitals (which in
the \textsc{Conquest} scheme are referred to as ``support functions'')
must be represented in terms of basis functions.  The choice of basis
function is a well-known problem in electronic structure; in this
context, the two extremes are those of small basis size and simplicity
(solutions of the atomic Schr\"odinger equation) and large basis size
and systematicity (allowing an increase in accuracy).  In the former
group are numerical atomic
orbitals\cite{sankey89,Kenny2000,Junq2001,soler02,Ozaki2003,Ozaki2004}
while in the latter are spherical waves\cite{haynes99a,gan01},
periodic sinc functions\cite{Mostofi2002,Mostofi2003},
wavelets\cite{arias99,Goedecker2005}, numerical
representation on a grid\cite{fattebert00} and blips
functions\cite{Hernandez1997}.  Recently, there has been work showing
that it is possible to define a systematically convergent set of
numerical orbitals (with respect to number of orbitals per angular
momentum channel)\cite{Ozaki2003,Ozaki2004}, and that numerical
orbitals can be fitted systematically to results of plane wave
calculations\cite{Kenny2000}.  The elements of the matrix $K_{i \alpha
  , j \beta}$ are then the elements of the density matrix in the
(generally non-orthogonal) representation of the $\{ \phi_{i \alpha}
\}$. To achieve $\mathcal{O}(N)$, the $\phi_{i \alpha} ( {\bf r} )$
are required to be non-zero only within finite regions, which can be
chosen as spheres of radius $R_{\rm reg}$, and $K_{i \alpha , j
  \beta}$ is also subject to a spatial cut-off. The ground state is
then found by minimising $E_{\rm tot}$ with respect to the orbitals
$\phi_{i \alpha} ( {\bf r} )$, if the basis set allows this, and
with respect to the $K_{i \alpha , j \beta}$, subject to idempotency
and fixed electron
number~\cite{hernandez96,hernandez95,bowler02}.  There are many
ways of finding the ground state density matrix, $K_{i \alpha , j
  \beta}$, including the auxiliary density matrix
method\cite{li93,Nunes1994}, penalty functional
methods\cite{kohn96,haynes99}, the Fermi operator expansion
(FOE)\cite{Goedecker1994}, bond-order potential
(BOP)\cite{Pettifor1989,Aoki1993,ozaki01}, and the constrained search
formalism\cite{ordejon93,mauri93,kim95}.  Many of these have
been compared for different materials\cite{bowler97,daniels97} and are
described in an extensive review article\cite{goedecker99}.

\section{Overview Of Conquest Methodology}
\label{sec:summ-conq-techn}

The practical implementation of any $\mathcal{O}(N)$ scheme raises two
important questions initially: first, which basis set to choose for
representing the localised orbitals (the most common choices were
mentioned above); and second, how to find the ground state density
matrix within the requirement of idempotency (again, several of the
more common schemes were mentioned above).  At present, in
\textsc{Conquest}, two basis sets are available: the B-splines, or
blips, which can be systematically converged to a given plane wave
result; and pseudo-atomic orbitals, which are economical and very
often give accurate results.  We envisage that we will implement other
options as required.  The main emphasis in the present paper is on
pseudo-atomic orbitals.


As mentioned above, there are various techniques in use for finding
the ground-state density matrix.  The present implementation in
\textsc{Conquest}\cite{Bowler1999} is a combination of the techniques of Li, Nunes and
Vanderbilt (LNV)\cite{li93} and Palser and
Manolopoulos~\cite{palser98}, both of which are closely related to
McWeeny's `purification' scheme~\cite{mcweeny60}. In the LNV
technique, the density matrix $K$ is represented in terms of an
`auxiliary' density matrix $L$ as:
\begin{equation}
K = 3 L S L - 2 L S L S L \; ,
\label{eq:LNV_DM}
\end{equation}
where $S$ is the overlap matrix for the support functions or localised
orbitals: $S_{\lambda \mu} = \langle \phi_\lambda | \phi_\mu \rangle$.
This scheme enforces `weak' idempotency~\cite{mauri93} (meaning that
all eigenvalues are in the interval $[ 0 , 1 ]$) rather than strict
idempotency.  If the total energy is minimised with respect to $L$
(writing $E_{\mathrm{GS}} = 2\mathrm{Tr}[KS]$ for the band energy),
this scheme automatically drives $K$ towards idempotency.  In order to
ensure $\mathcal{O}(N)$ scaling, a spatial cut-off is imposed on the
$L$-matrix, so that $L_{\lambda \mu} = 0$ when the distance between
the centres of the support-functions $\phi_\lambda$ and $\phi_\mu$
exceeds a chosen cut-off $R_L$.  Other methods for finding the
ground-state density matrix could be implemented with little effort.

As well as operating in the $\mathcal{O}(N)$ mode, \textsc{Conquest}
can find the ground state directly by diagonalisation, using the
\textsc{ScaLapack} package, which allows efficient parallelisation of
the diagonalisation.  Since it scales as $\mathcal{O}(N^3)$, this will
only be approrpriate for relatively small systems, but it provides an
important tool both for testing the outer parts of the ground-state search
(described below) and for exploring the convergence of the
$\mathcal{O}(N)$ algorithm with the cut-off on the $L$-matrix.

The search for the ground-state is organised into three loops.  In the
innermost loop, the support functions and electron density are fixed
and the ground-state density matrix is found, either by varying $L$ or
by diagonalisation. In the middle loop, self-consistency is achieved
by systematically reducing the electron-density residual, i.e. the
difference between the input and output density in a given
self-consistency cycle~\cite{johnson88}.  In the outer loop, the
energy is minimised with respect to the support functions,
$\phi_\lambda$. This organisation corresponds to a hierarchy of
approximations: when the inner loop alone is used, we get the scheme
known as non-self-consistent {\em ab initio} tight binding (NSC-AITB),
which is a form of the Harris-Foulkes
approximation~\cite{harris85,foulkes89,sankey89,horsfield00}; when the
inner two loops are used, we get self-consistent {\em ab initio} tight
binding (SC-AITB); finally, if all loops are used, we have full {\em
  ab initio}. In this last case, we recover the exact DFT ground state
as the region radius $R_{\rm reg}$ and the $L$-matrix cut-off $R_L$
are increased.  For non-metallic systems, the evidence so far is that
accurate approximations to the ground state are obtained with quite
modest values of the cut-offs~\cite{hernandez96,soler02}.  

For calculations at the level of full \textit{ab initio} accuracy, the
convergence of the outer loop (optimising the support functions with 
respect to their basis functions) is well-conditioned provided
appropriate pre-conditioning measures are taken; these have been
discussed both for blips in the context of
\textsc{Conquest}\cite{bowler1998a,Gillan1998a} and for psinc functions in the
context of \textsc{onetep}\cite{Mostofi2003}.  We note that
\textsc{Conquest} can be run in a mode analagous to \textsc{siesta},
where pseudo-atomic orbitals are used and no optimisation is
performed; in this case, the outer loop is not performed.

We have recently found that the self-consistency search (the middle
loop described above) can be accelerated by use of the Kerker
preconditioning.  This idea, which is well-known in the plane-wave
community, removes long wavelength changes in the charge density
during mixing.  It is applied in reciprocal space, as a prefactor:
\begin{equation}
  \label{eq:2}
  f(q) = \frac{q^2}{q^2 + q^2_0}
\end{equation}
Then the charge is mixed using a Pulay or Broyden (or related)
scheme\cite{johnson88} with the prefactor applied to the residual or
output charge after transformation to reciprocal space. The mixing
includes a parameter, A, which determines how aggressive the mixing is
(with the input charge density for iteration $n+1$ given by
$\rho^{n+1}_{in} = \rho^n_{in} + A f(q) R_n$, with $R_n$ the residual
from iteration $n$).  The efficacy of this preconditioning is explored,
both for exact diagonalisation and linear scaling, in
Section~\ref{sec:self-consistency}.

While performing the search for self-consistency, we must monitor the
residual.  We define the following dimensionless parameter which is
used to monitor the search:
\begin{eqnarray}
  \label{eq:4}
  d &=& \frac{\langle \left| R(\mathbf{r}) \right|^2
    \rangle^{1/2}}{\bar{\rho}},\\
  \label{eq:3}
  \langle \left| R(\mathbf{r}) \right|^2 \rangle &=& \frac{1}{V} \int
  \mathrm{d}\mathbf{r} \left| R(\mathbf{r}) \right|^2,
\end{eqnarray}
where $V$ is the simulation cell volume and we use the usual
definition of residual, $R(\mathbf{r}) = \rho_{\mathrm{out}}
(\mathbf{r}) - \rho_{\mathrm{in}}(\mathbf{r})$, the difference between
the output and input charge densities.  The quantity $d$ is
then the RMS value of $R(\mathbf{r})$ normalised by dividing by the
\emph{average} charge density in the system, $\bar{\rho}$.  Note that,
for systems containing large amounts of vacuum, the criterion for
convergence will need to be altered when compared to bulk-like
environments.  This criterion may be coupled with a monitor on the
largest value of residual on an individual grid point $\mathbf{r}_l$,
$R_{\mathrm{max}} = \max_l \left|R(\mathbf{r}_l)\right|$

The scheme we have outlined is closely related to the methods used in
\textsc{siesta}~\cite{ordejon96,soler02}, OpenMX\cite{Ozaki2003} and
\textsc{onetep}\cite{Skylaris2005}.  The main differences are: (i) the basis
sets chosen (\textsc{siesta} uses fixed PAOs, while OpenMX uses
optimized orbitals and \textsc{onetep} psinc functions); (ii) the method of
finding the ground state density matrix (Siesta uses the constrained
search technique\cite{mauri93,kim95,ordejon93}, OpenMX the
divide-and-conquer\cite{Yang1991} or BOP\cite{ozaki01} and \textsc{onetep}
either penalty functional\cite{kohn96,haynes99a} or
LNV\cite{li93}); (iii)~the technique of `neutral-atom
potentials'~\cite{ordejon96,soler02}, used by \textsc{siesta} and
OpenMX, which allows calculation of matrix elements to be performed
very efficiently for localised, atomic-like basis sets.

The principle of near-sightedness (i.e. spatial locality of electronic
structure), means that sparse matrix multiplications and other
operations are restricted to a finite area of space, leading to a
natural route to parallel decomposition of the problem.
\textsc{Conquest} was written from the outset as parallel code, and a
large part of the development effort has been concerned with
techniques for achieving good parallel scaling. The parallelisation
techniques have been described in detail
elsewhere~\cite{goringe97,bowler01,bowler02}, so we give only a brief
summary. There are three main types of operation that must be
carefully distributed across processors:

\begin{itemize}
\item
the storage and manipulation of localised orbitals, e.g. the
calculation of $\phi_\lambda ( {\bf r} )$ on the integration grid
starting from blip- or PAO-coefficients, and the calculation of
the derivatives of $E_{\rm tot}$ with respect to these coefficients,
which are needed for the ground-state search;

\item
the storage and manipulation of elements of the various
matrices ($H$, $S$, $K$, $L$, etc...);

\item
the calculation of matrix elements by summation over
domains of points on the integration grid, or by analytic operations
(for certain integrals involving PAOs and blips).
\end{itemize}

Efficient parallelisation of these operations, and the 
elimination of unnecessary communication between processors,
depend heavily on the organisation of both atoms and grid points
into small compact sets, which are assigned to processors~\cite{bowler01}.
When the code runs in $\mathcal{O}(N)$ mode, matrix multiplication takes
a large part of the computer effort, and we have developed
parallel multiplication techniques~\cite{bowler01} that exploit the specific patterns
of sparsity on which $\mathcal{O}(N)$ operation depends.

\section{Calculation of Ionic Forces}
\label{sec:ionic-forces}

In order to perform structural relaxation or molecular dynamics of
materials with an electronic structure technique, the algorithms for 
calculating the forces $\mathbf{F}_i$ on the ions must be the exact
derivatives of the total ground state energy, $E_{\mathrm{GS}}$, with
respect to the positions, $\mathbf{r}_i$, such that $\mathbf{F}_i =
-\nabla_i E_{\mathrm{GS}}$.  One of the advantages of DFT, within the
pseudopotential approximation, is that it is easy, in principle, to
achieve this relationship between the forces and the energy.  Since
the \textsc{Conquest} formalism allows the calculation of the total
energy at different levels of accuracy, some care is needed in the
formulation of the forces to develop a scheme that works at all levels
of this hierarchy.  It is also important to ensure that it works
equally well (and accurately) for both the diagonalisation and
$\mathcal{O}(N)$ modes of operation implemented in \textsc{Conquest}.
We have recently described these algorithms in
detail\cite{Miyazaki2004}, but we summarise them below for convenience.


We recall the Harris-Foulkes expression\cite{harris85,foulkes89} for
the total energy, which is often applied when self-consistency is not
sought, but which at self-consistency is identical to the
standard Kohn-Sham expression for total energy.  The expression is:

\begin{equation}
  \label{eq:1}
  E_{\mathrm{GS}} = E_{\mathrm{BS}} + \Delta E_{\mathrm{Har}} + \Delta
  E_{\mathrm{xc}} + E_\mathrm{C},
\end{equation}
with $E_\mathrm{C}$ the Coulomb energy between the ionic cores, and
the band-structure energy, the double-counting Hartree and
exchange-correlation energies defined as:

\begin{eqnarray}
E_\mathrm{BS} &=& 2 \sum_n f_n \epsilon_n \\
&=& 2\mathrm{Tr}[KH]\\
\Delta E_\mathrm{Har} & = & - \frac{1}{2} \int d \mathbf{r} \, n^\mathrm{in} ( \mathbf{r} )
V_\mathrm{Har}^\mathrm{in} ( \mathbf{r} ) \nonumber \\
\Delta E_\mathrm{xc} & = & \int d \mathbf{r} \, n^\mathrm{in} ( \mathbf{r} )
\left( \epsilon_\mathrm{xc} ( n^\mathrm{in} ( \mathbf{r} )) - \mu_\mathrm{xc} (
n^\mathrm{in} ( \mathbf{r} ) ) \right) \; .
\label{eqn:E_double_count}
\end{eqnarray}
Here, $n^\mathrm{in} ( \mathbf{r} )$ is the \emph{input} charge
density used (normally a superposition of atomic charge densities if a
non-self-consistent scheme is used, or the self-consistent charge
density if self-consistency is used).  This expression is very useful
when comparing forces at different levels of approximation.

At the empirical TB level, the ionic force is a sum of the
band-structure part ${\bf F}_i^{\rm BS}$ and the pair-potential
part ${\bf F}_i^{\rm pair}$, the former being given by~\cite{sankey89}:
\begin{equation}
{\bf F}_i^{\rm BS} = - 2 {\rm Tr} \; [ K \nabla_i H -
J \nabla_i S ] \; ,
\end{equation}
where $K$ and $J$ are the density matrix and energy matrix
respectively~\cite{sankey89}. It is readily shown that in the $\mathcal{O}(N)$ scheme
of LNV, and in some other $\mathcal{O}(N)$ schemes, the same formula
for ${\bf F}_i^{\rm BS}$ is the exact derivative of the
$\mathcal{O}(N)$ total energy. In the LNV scheme, $K$ is given by
eqn~(\ref{eq:LNV_DM}), and $J$ by:
\begin{equation}
J = - 3 L H L + 2 L S L H L + 2 L H L S L \; .
\end{equation}

In NSC-AITB (Harris-Foulkes), the forces can be written
in two equivalent ways. The way that corresponds
most closely to empirical TB is:
\begin{equation}
{\bf F}_i = {\bf F}_i^{\rm BS} + {\bf F}_i^{\Delta {\rm Har}} +
{\bf F}_i^{\Delta {\rm xc}} + {\bf F}_i^{\rm ion} \; ,
\end{equation}
where ${\bf F}_i^{\rm BS}$ is given by exactly the same formula as
in empirical TB. The contributions ${\bf F}_i^{\Delta {\rm Har}}$ and 
${\bf F}_i^{\Delta {\rm xc}}$, which arise from the double-counting
Hartree and exchange-correlation parts of the NSC-AITB total energy,
have been discussed elsewhere~\cite{sankey89}. The final term ${\bf F}_i^{\rm ion}$
come from the ion-ion Coulomb energy. This way of writing
${\bf F}_i$ expresses the well-known relationship between
NSC-AITB and empirical TB that in the latter the pair term
represents the sum of the three contributions
$\Delta {\rm Har} + \Delta {\rm xc} + {\rm ion-ion}$.
The alternative, and exactly equivalent, way of writing
${\bf F}_i$ in NSC-AITB is:
\begin{equation}
{\bf F}_i = {\bf F}_i^{\rm ps} + {\bf F}_i^{\rm Pulay} +
{\bf F}_i^{\rm NSC} + {\bf F}_i^{\rm ion} \; .
\label{eq:NSC_AITB_2}
\end{equation}
Here, ${\bf F}_i^{\rm ps}$ is the ``Hellmann-Feynman'' force
exerted by the valence electrons on the ion cores;
${\bf F}_i^{\rm Pulay}$ is the Pulay force that arises in
any method where the basis set depends on ionic positions;
${\bf F}_i^{\rm NSC}$ is a force contribution associated with
non-self-consistency, and is expressed in terms of the
difference between output and input electron densities;
${\bf F}_i^{\rm ion}$, as before,
is the ion-ion Coulomb force. Exactly the same formulas
represent the exact derivative of $E_{\rm tot}$ in both diagonalisation
and $\mathcal{O}(N)$ modes.

In both SC-AITB and full AI, the force formula is:
\begin{equation}
{\bf F}_i = {\bf F}_i^{\rm ps} + {\bf F}_i^{\rm Pulay} +
{\bf F}_i^{\rm ion} \; ,
\end{equation}
which differs from the second version of the
NSC-AITB formula eqn~(\ref{eq:NSC_AITB_2}) only by the absence of the non-self-consistent
contribution ${\bf F}_i^{\rm NSC}$, as expected.

The above hierarchy of force formulas has been implemented
in \textsc{Conquest}, and extensive tests have ensured
that the total energy and the forces are exactly consistent
within rounding-error precision\cite{Miyazaki2004}.

\section{Illustrative Results}
\label{sec:illustrative-results}

We have already demonstrated the ability of Conquest to address
non-trivial systems by relaxing the Si(001) surface with a variety of
basis sets, and comparing the results to geometries obtained with both Siesta and
VASP\cite{Miyazaki2004}.  In this section, we will present further
results which show that \textsc{Conquest} is now ready for application
to real-world scientific problems.

We start with the problem of self-consistency, and demonstrate both that
the Kerker preconditioning technique described in
Section~\ref{sec:summ-conq-techn} above is extremely useful and that
it does not degrade the linear scaling stability.  We then present
selected data from a study of the three-dimensional ``huts'' which
evolve to relieve strain during heteroepitaxial growth of Ge on
Si(001).

\subsection{Self-consistency}
\label{sec:self-consistency}

The search for self-consistency between the charge density and
potential can still sometimes be problematic in standard DFT codes:
even with the most sophisticated, convergence to the ground state
is not completely guaranteed.  In this section, we present the results of
tests on three different systems, demonstrating the effects of varying
the mixing parameter, A, and the preconditioning wavevector, q$_0$ in
the Kerker preconditioning, as discussed in
Section~\ref{sec:summ-conq-techn}.  The results will be both for exact
diagonalisation and linear scaling ground state search methodologies,
though we reserve the linear scaling results for the most challenging
problem, presented last.  We present the raw data (change of residual
with iteration) for selected cases at the end of the section, along
with data on the effect of region radius on convergence rate.

The parameters used in modelling the systems are as follows:
\textsc{Conquest} was operating with a PAO basis set at the
single-zeta level (with four orbitals per atom) following the
generation scheme in \textsc{siesta}, with cutoffs of 4.88 and 6.12
bohr on the s- and p-orbitals respectively (corresponding to a shift
of 250 meV on-site).  There was no outer loop (as discussed in
Sec.~\ref{sec:summ-conq-techn} above).  The criterion applied for
reaching self-consistency was $d$=0.01 for bulk silicon (where $d$ is
defined in Eq.~(\ref{eq:4}) above). This value of $d$ is perfectly
adequate for total energy calculations though may be a little loose
for accurate molecular dynamics; for this purpose it is ideal as it is
the early stages of the self-consistency search which are non-linear.
For both the silicon surface and the cluster, the \emph{average}
electron density in the system is smaller owing to the vacuum present;
in both cases we used larger values of $d$ so that the effective
convergence was the same (for Si(001) we set $d$=0.015 and for the
silicon cluster $d$=0.039).  In all cases, the starting charge density
was a linear superposition of atomic charge densities generated from
the PAO basis functions.

\begin{vchfigure}
  \centering
  \includegraphics[clip,width=0.5\textwidth]{pssb200541386_Fig1}
  \vchcaption{Convergence to self-consistency with Kerker
    preconditioning wavevector and mixing parameter, A, for 512 atom cell
    of bulk silicon.  Any value of 50 iterations indicates a lack of
    convergence for that parameter.}
  \label{fig:BulkSi_SC}  
\end{vchfigure}

We begin with the simplest possible system: bulk silicon.  The results
for a unit cell containing 512 atoms are given in
Fig.~\ref{fig:BulkSi_SC}, purely for diagonalisation at the gamma
point.  As might be expected for such a simple system, there is no
problem reaching the ground state, and in fact the Kerker
preconditioning is ineffective: at best it has no effect, and it often
slows down the iterations.

\begin{vchfigure}
  \centering
  \includegraphics[clip,width=0.5\textwidth]{pssb200541386_Fig2}
  \vchcaption{Convergence to self-consistency with Kerker
    preconditioning wavevector and mixing parameter, A, for 192 atom cell
    of Si(001).  Any value of 50 iterations indicates a lack of
    convergence for that parameter.}
  \label{fig:Si001_SC}  
\end{vchfigure}

We now move from a three-dimensional system to the Si(001) surface.
The results for a slab containing 192 atoms (the unit cell was one
dimer row long, eight dimers long, and 12 layers deep with a surface
on both sides) are given in Fig.~\ref{fig:Si001_SC}.  The Kerker
preconditioning is vital here, with the value of q$_0$ determining
stability during the search.

\begin{vchfigure}
  \centering
  \includegraphics[clip,width=0.5\textwidth]{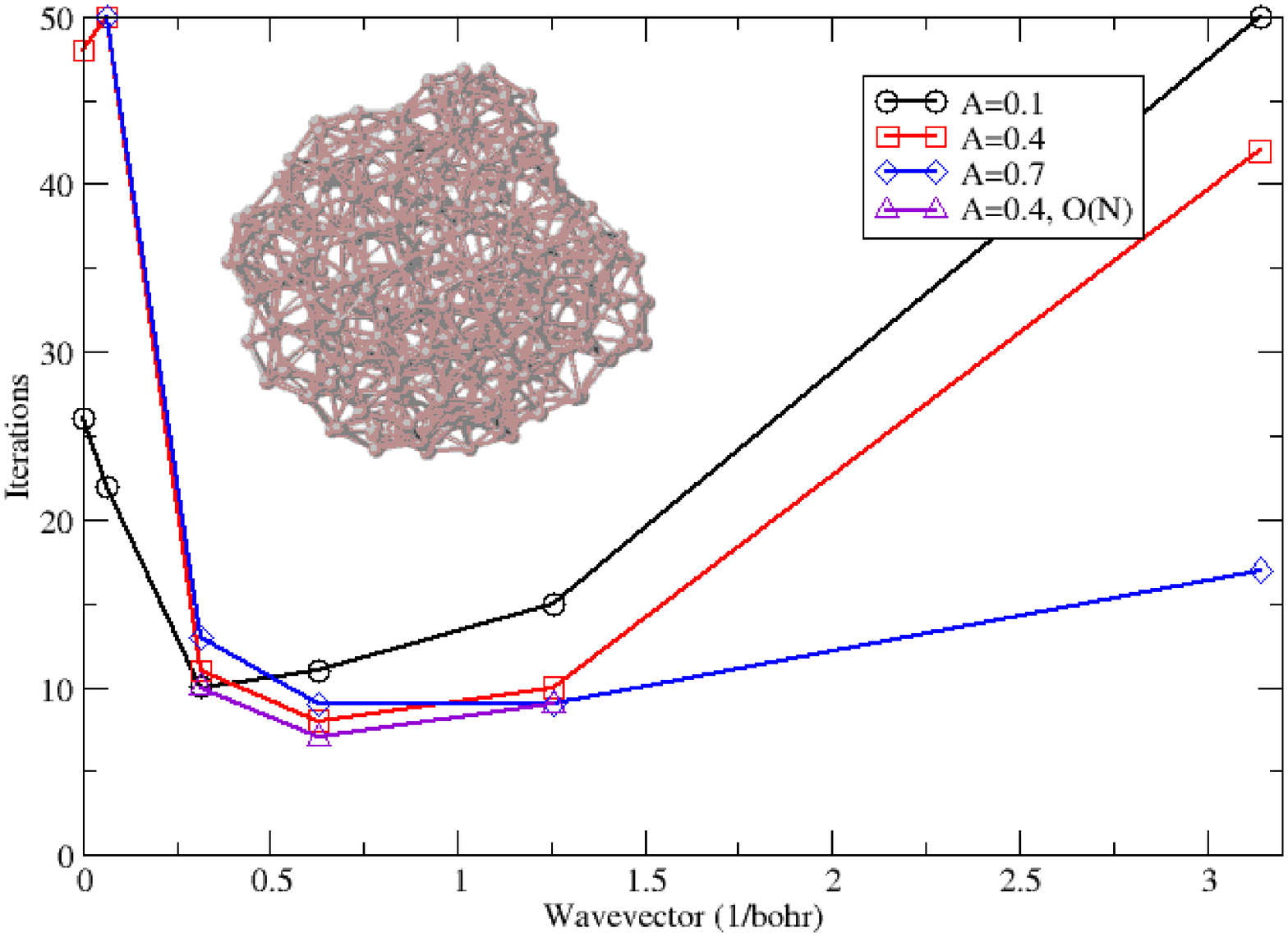}
  \vchcaption{Convergence to self-consistency with Kerker
    preconditioning wavevector and mixing parameter, A, for amorphous,
    metallic 343 atom cluster of silicon.  Any value of 50 iterations
    indicates a lack of convergence for that parameter.}
  \label{fig:SiCluster_SC}  
\end{vchfigure}

Finally we turn to an amorphous cluster of 343 silicon atoms, whose
structure was optimised by a minima-hopping
procedure\cite{Goedecker2004}.  A projection of the structure of the
cluster is shown in Fig.~\ref{fig:SiCluster_SC}, along with the
convergence results.  The results for exact diagonalisation have been
reproduced with linear scaling to test the effect of Kerker mixing on
the stability of the algorithm (shown with downward triangles).  As is
clear from the plot, the use of linear scaling for this cluster is not
affected at all by the choice of self-consistency algorithm, and
convergence \emph{to self-consistency} is equally fast with either
method.  We note that the cluster is metallic (or at least that there
is partial occupation of a number of levels near the Fermi level),
leading to rather slow convergence of the density matrix minimisation
(though we note that the cutoff on the density matrix was 7\,\AA,
which is not large enough to provide convergence to the
diagonalisation result).  Nevertheless, the self-consistent ground
state is achieved without significant difficulty.

We note that the optimal values of $q_0$ found are in good agreement
with those previously published, though there is little quantitative data
available beyond Ref.~\cite{Kresse1996a}.  In that publication, it was
stated that values of $q_0$ in the range 0.5---2.0 \AA$^{-1}$ yielded
little change in the convergence rate.  The factor to convert between
bohr$^{-1}$ and \AA$^{-1}$ which is required in order to compare this range to
the data in Figs.~\ref{fig:BulkSi_SC}--\ref{fig:SiCluster_SC} is
1.8897 (simply one bohr$^{-1}$).  Then we see that, excluding
bulk silicon, the acceptable range of 0.20--0.65 bohr$^{-1}$ for
Si(001) and 0.30--1.25 bohr$^{-1}$ for the cluster convert to
0.38--1.23~\AA$^{-1}$ and 0.57--2.36~\AA$^{-1}$, in very good
agreement.  We also note that the problem of slow convergence for
overly small mixing parameter A noted\cite{Kresse1996a} is seen in
Fig.~\ref{fig:Si001_SC}. 

\begin{vchfigure}
  \centering
  \includegraphics[clip,width=0.5\textwidth]{pssb200541386_Fig4}
  \vchcaption{Residual, defined in Eq.~(\protect\ref{eq:4}), during
    self-consistency search.  The system considered is the amorphous
    cluster shown in Fig.~\ref{fig:SiCluster_SC}, and different
    conditions are considered for the simulation as discussed in the
    main text.}
  \label{fig:RawData}  
\end{vchfigure}

The raw data, that is value of $d$ per iteration, is shown in
Fig.~\ref{fig:RawData} for the 343 atom cluster.  For $q_0$=0.2 and
A=0.1, we show convergence for the single zeta PAO basis (circles) and
an \emph{unconverged} blip basis (squares).  This shows that the
convergence rate is only mildly dependent on choice of basis, or
indeed on the convergence of the outer loop (since the blips chosen
were far from optimal).  Similarly, for $q_0$=0.1 and A=0.4, we show
convergence for both exact diagonalisation (diamonds) and linear
scaling (triangles) solvers .  Again, the rate of convergence is only
mildly changed by this change of methodology.  These results are
illustrative of the general behaviour of the minimisation: the
self-consistency scheme described is unaffected by the details of the
inner or outer loops.

\begin{vchfigure}
  \centering
  \includegraphics[clip,width=0.5\textwidth]{pssb200541386_Fig5}
  \vchcaption{Residual, defined in Eq.~(\protect\ref{eq:4}), during
    self-consistency search for different region radii.  The system
    considered is the amorphous cluster shown in
    Fig.~\ref{fig:SiCluster_SC}, with the convergence used for that
    figure, and Kerker values A=0.4, $q_0$=0.628 bohr$^{-1}$.  Details
    of basis sets are given in the text.}
  \label{fig:RawDataRc}  
\end{vchfigure}

Finally, we present further a further plot of the residual per
iteration for the 343 atom Si cluster for different values of the
region radius in Fig.~\ref{fig:RawDataRc}.  For this test, we have
again used single-zeta PAOs, with different techniques for generation.
We used both the \textsc{siesta} energy shift (yielding radii of
5.67/7.11 bohr for s/p with 100 meV, 5.26/6.43 bohr for 200 meV and
4.88/6.12 bohr for 300 meV) and fixed radii for both s and p channels
(of 6.0 bohr, 6.6 bohr, 7.0 bohr and 8.0 bohr).  We see that there is
only a very small dependence of convergence on region radius. 

\subsection{Hut clusters of Ge on Si(001)}
\label{sec:hut-clusters-ge}

\begin{vchtable}
  \vchcaption{Surface energy for Ge(105) calculated using exact
    diagonalisation (labelled diag) and linear scaling with different
    density matrix cutoff distances.}\label{tab:Ge105}
  \begin{tabular}{llll}
    $R_L$(bohr) & E$_{\mathrm{tot}}$ (Ha) & F$_{\mathrm{max}}$ (Ha/bohr)& E$_{\mathrm{surf}}$(eV/A$^2$) \\
    \hline
    15.4  & -484.7635 & 0.0027 & 0.0801\\  
    20.4  & -485.0438 & 0.0014 & 0.0753\\  
    25.4  & -485.1420 & 0.0007 & 0.0752\\  
    30.4  & -485.1811 & 0.0004 & 0.0755\\  
    diag  & -485.2048 & 0.0001 & 0.0765    
  \end{tabular}    
\end{vchtable}

We are currently investigating the stability of three-dimensional
islands of Ge on Si(001) during strained heteroepitaxial
growth\cite{Stangl2004}, following our earlier studies of strained
layer growth using conventional DFT\cite{Oviedo2002} and linear
scaling tight binding\cite{Li2003}.  After approximately three
monolayers of Ge has been deposited (though this varies with the
system and growth conditions), small mounds with well-defined facets
begin to appear.  These have facets made of the Ge(105) surface, and
allow significant strain relaxation.  A first step in understanding
these huts is to characterise the Ge(105) surface\cite{Fujikawa2002},
which we have done with \textsc{Conquest}, in particular concentrating
on the density matrix range required for accurate calculations.  We
present surface energy and maximum forces calculated using a minimal
basis at the NSC-AITB level for a Ge(105) surface in
Table~\ref{sec:hut-clusters-ge}. (We note that the structure modelled
here is a previously unreported insulating structure which will be
presented in more detail elsewhere.) The surface energy was calculated
using systems with eight layers and ten layers (100 and 124 atoms
respectively) The system was relaxed using exact
diagonalisation (with a $4\times 5\times 1$ Monkhorst-Pack mesh) and
this structure was used in the linear scaling calculations.  We see
that the maximum force is converged at a density matrix range of 20 or
25 bohr, but that the surface energy is more slowly convergent.
Nevertheless, this is a strong indication that reliable results can be
obtained using linear scaling\footnote{In fact, on converging the
  $\mathcal{O}(N)$ results it was found that the exact diagonalisation
  results were insufficiently converged with respect to the k-point
  mesh, and the mesh had to be increased.}.

\begin{vchfigure}
  \centering
  \includegraphics[clip,width=0.5\textwidth]{pssb200541386_Fig6}
  \vchcaption{Convergence during minimisation of energy with respect
    to density matrix elements (innermost loop of ground state search)
    for 23,000 atom Ge hut cluster on Si(001) system.}
  \label{fig:LargeHut_DMM}  
\end{vchfigure}

We have considered a number of different hut cluster systems, with
different sizes and spacings on the substrate.  The largest system
which we have prepared so far contains approximately 23,000 atoms
(including substrate, wetting layer of Ge and the hut cluster).  We
show the residual of the density matrix against iteration number in
Fig.~\ref{fig:LargeHut_DMM}, which demonstrates both that convergence
to the ground state can be easily achieved for a system this large,
and that the number of iterations needed to reach the ground state is
independent of system size (which is an important consideration for
this type of code).

\begin{vchfigure}
  \centering
  \includegraphics[clip,width=0.5\textwidth]{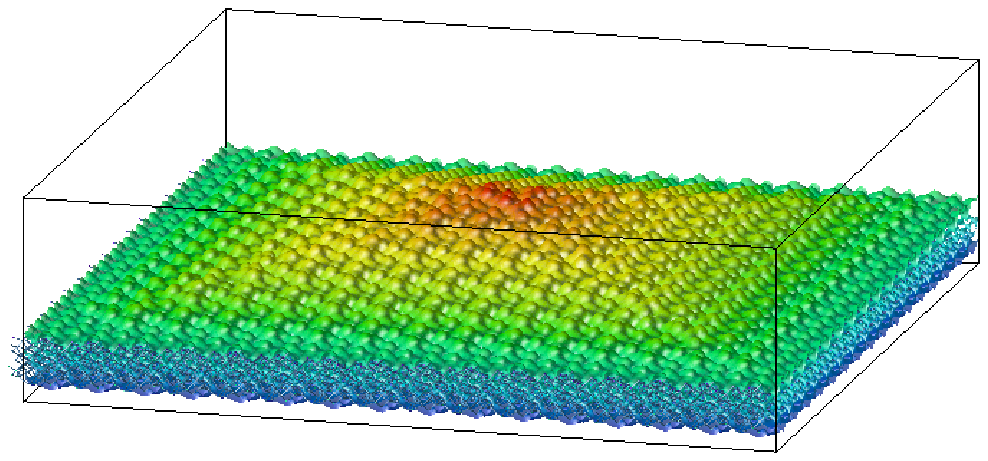}
  \vchcaption{Charge density from 23,000 atom Ge hut cluster on
    Si(001) system.  The density was sampled at one grid point in four
    in each direction to create a manageable data set.  Colour
    indicates height; isosurface is equivalent to $\sim$0.3 electrons
    per cubic \AA ngstrom.}
  \label{fig:LargeHut_Charge}  
\end{vchfigure}

Finally, we present a preliminary charge density for this system in
Fig.~\ref{fig:LargeHut_Charge}, where the colour indicates height in
the unit cell.  The isosurface was formed by sampling one grid point
in four in each direction, to make the data set manageable.  The
charge density shows both the scale of the simulation and the dimers
on the facets of the hut, as well as the characteristic buckling
associated both with the Si(001) surface and the Ge(105) faces of the
hut cluster.

\section{Discussion And Conclusions}
\label{sec:disc-concl}

We have presented an overview of the methodology used in the
\textsc{Conquest} code, and shown that realistic calculations on
complex, scientifically interesting systems are now possible.  In
particular, we have outlined the arrangement of the code to allow
calculations at different levels of precision (from minimal-basis AITB
through to full \textit{ab initio}) and presented details of how
forces can be calculated consistently at all levels of precision.  We
have shown details of convergence to self-consistency for various
different systems, and outlined early results from our study of
three-dimensional ``hut'' clusters resulting from heteroepitaxial
growth of Ge on Si(001).

We want to note that there are other $\mathcal{O}(N)$ techniques
within the DFT arena, notably Siesta\cite{soler02},
OpenMX\cite{Ozaki2003,Ozaki2004} and \textsc{onetep}\cite{Skylaris2005}.
However, the ideas of linear scaling are not limited to DFT, and have been
used in $\mathcal{O}(N)$ Hartree-Fock
methods\cite{challacombe99,Ayala1999,Lee2000}.  Very recently, some of
these ideas have been applied to Quantum Monte Carlo
techniques\cite{williamson01}, and it has been shown that localised
orbitals with blip-function basis sets are capable of giving a major
speed-up to Quantum Monte Carlo calculations\cite{Alfe2004}.  There
are clear signs that $\mathcal{O}(N)$ methods are realising their
early promise.


\begin{acknowledgement}
  The \textsc{Conquest} project is partially supported by ACT-JST.
  DRB is supported by a Royal Society University Research Fellowship,
  and RC by an EPSRC studentship.  This study was performed through
  Special Coordination Funds for Promoting Science and Technology from
  the MEXT, Japan.
\end{acknowledgement}

\end{document}